# INFLUENCE OF ELECTRIC CURRENT ON DENSIFICATION OF CONDUCTIVE POWDERS IN SPARK PLASMA SINTERING

Geuntak Lee[1,2], Charles Manière[1], Joanna McKittrick[2], Eugene A. Olevsky[1,3*]

[1]Mechanical Engineering, San Diego State University, San Diego, USA
[2]Mechanical and Aerospace Engineering, University of California, San Diego, La Jolla, USA
[3]NanoEngineering, University of California, San Diego, La Jolla, USA

## ABSTRACT

The 3 heating modes are utilized to make ZrN powders have 3 different levels of the electric current density at the same temperature during spark plasma sintering (SPS). The constitutive equation of sintering for SPS is applied to the experimental porosity evolution of ZrN from three SPS modes, and this showed high electric current density increase the electric current assisted deformability value of ZrN pellets, resulting in a reduction of the flow stress. The electric current flow enhances the dislocation motion, which was experimentally proved and analyzed by modified Williamson-Hall equation applying to X-ray diffraction results, and the mechanical strength test of ZrN pellets.

[*] Corresponding author: e-mail: eolevsky@mail.sdsu.edu

Densification of the powders like metal and ceramic can be occurred by three mechanisms: sintering, particle rearrangement, and plastic deformation [1]. There is much debate on if electric current affects the densification behaviors of powders during spark plasma sintering (SPS). In other words, this comes to end the basic question that if electron flow influences the densification of the powders except for Joule heating effect. Electro plastic effect (EPE) can give some hint to this question.

Electroplastic theory shows that the electron flow helps the dislocation motion through lattice structure, resulting in the reduction of flow stress required for the deformation of solid material (meaning fully dense material) [2, 3]. This effect can be caused by two possible mechanisms. At first, electrons collision and scattering to defects generate the local increase in the energy of atomic vibration near the dislocation core and the rate of diffusion [4, 5]. Secondly, the electron transfers extra momentum directly to obstacles such as impurities to be removed [3, 6]. Both mechanisms explain the increase of the formability of the material and reduction of the deformation energy [7]. However, some researchers indicated that the electron wind effect is not enough to affect the dislocation activities compared with first mechanism [4, 8, 9].

Back to SPS, EPE can easily explain the large plastic deformation of the particle itself under electric current [10]: the electron flow can enhance the dislocation mobility so plastic deformation of the non-porous particle will be severed for metal at low temperature and ceramic at high temperature, which aids the densification. Additionally, the local thermal energy at defects by electron flow help the diffusion of atoms, enhancing sintering.

Also, SPS is pressure-assisted densification process. Interestingly, by inference from electroplastic theory and experiment [5, 11], the pressure can give synergetic effect (other than particle rearrangement [12]) to densification when electric current applied. Siopis et al. [11], showed that cold-worked copper specimens which generating dislocations presented lower flow stress in electroplastic deformation process During SPS, applied pressure increase the dislocation density near neck area [13], so flow stress is decreased more with higher pressure under SPS condition, enhancing the total densification rate.

In the previous report [14], we showed the new constitutive equation for SPS with consideration of the electroplastic effect. Aside from thermal effect (Joule heating, heating rate or temperature gradient etc. [15]), the electric current effect term, $A_{ECAD}$ (electric current assisted deformability), in this equation explains how electric current density affect the formability or high-temperature ductility (ceramic) of material during the SPS.

In this study, the effect of electric current flow on material's deformability ($A_{ECAD}$) and dislocation density during ZrN densification process were elucidated by special 3 SPS modes experiments enabling the deconvolution of the electric current effects from the thermal contribution.

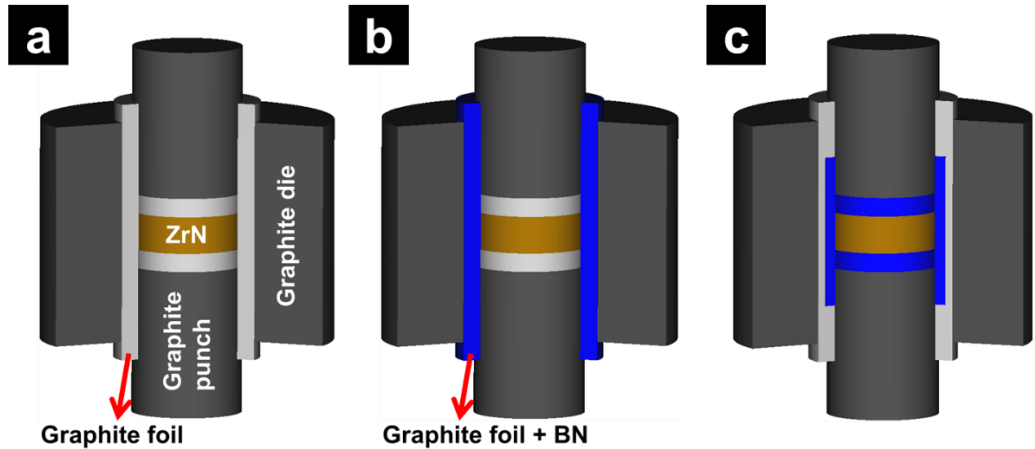

Fig. 1. Schematics of SPS setup enabling the different current density flow into ZrN powders. Black, gray, gold and blue color indicate the graphite component, graphite paper, ZrN, and boron nitride respectively. (a) Normal J mode, (b) High J mode and (c) Insulation mode.

ZrN powders (FCC, Fm3m) produced by Sigma-Aldrich with an average particle size of 2 μm were employed. The sintering experiments were carried out using the SPS device manufactured by SPS Syntex Inc., (Dr. Sinter SPSS-515, USA) with a graphite (EDM-4, Poco Graphite, Inc., TX, USA)) tooling setup. As shown in Fig. 1, boron nitride (BN) was coated on the graphite paper (0.15 mm-thick) in order to manipulate the current path in the SPS setup. The three SPS modes include "normal J (electric current density) mode" where the electric current can pass both in the sample and the die (Fig.1(a)), the "high J mode" where all the current passes through the sample and not to the die (Fig.1(b)) and, finally, "insulation mode" where no electric current gets across the sample (Fig.1(c)). For all the 3 modes, the same SPS conditions were used: 1600 °C, 60 MPa, 100 °C/min and 60 min holding time. The relative density was obtained by the Archimedes' immersion method. The fractured specimens were analyzed by the scanning electron microscopy (SEM), (FEI Quanta 450, USA).

For the deconvolution of electric current effect from the thermal effect, the exact temperature of the sample in SPS should be measured. Due to the high sintering temperature (1600 °C), the pyrometer was preferred as a temperature detecting and PID (proportional–integral–derivative) controlling device [16]. For all modes, ZrN powder was pre-consolidated by SPS until it had enough binding strength. The measured density after pre-consolidation was 65 %. Without removing the ZrN sample from the die in the pre-consolidation experiment, 0.5 mm depth hole at the side of the ZrN sample was made through 2 mm diameter graphite die hole, so that the optical pyrometer directly measured the temperature focused on the side of the pre-consolidated ZrN samples. This method guarantees that the temperature of ZrN samples is the same in all 3 SPS modes experiments. To compensate the relatively low accuracy of the temperature measurement by the pyrometer, the additional k-type thermocouple detected the temperature in another hole of the same ZrN specimen. The temperature difference between them was obtained as follows:

$$T_{ZrN\_TC} = 1.121 \times T_{ZrN\_Pyro} - 76.330 \qquad (1)$$

Where $T_{ZrN\_Pyro}$ and $T_{\_ZrN\_TC}$ are the temperatures (in °C) measured for ZrN by the pyrometer and by the thermocouple, respectively. By this method, instead of 1600 °C obtained from a pyrometer, 1717 °C obtained by Eq. (1) was used for the following study.

Because the processing conditions and measured sample temperature were same for the 3 processing modes during SPS, the electric current is the only factor to affect the densification behavior of ZrN powders. The total electric current values recorded by the SPS system ($I_t$) are not identical to the amount of the electric current ($I_s$) passing through the samples during SPS. $I_s$ value is clear for the high J (100% of $I_t$) and the insulation modes (0% of $I_t$). However, for the normal J mode $I_s$ is not the same as $I_t$, and is mainly affected by the electrical and thermal properties of the graphite die set. An electro-thermal model based on the finite element code COMSOL$^{TM}$ has been developed within the framework of the present studies to estimate the $I_s$ during the normal J mode. The properties of the materials considered and the detailed boundary conditions of the FEM simulations have been previously described [14].

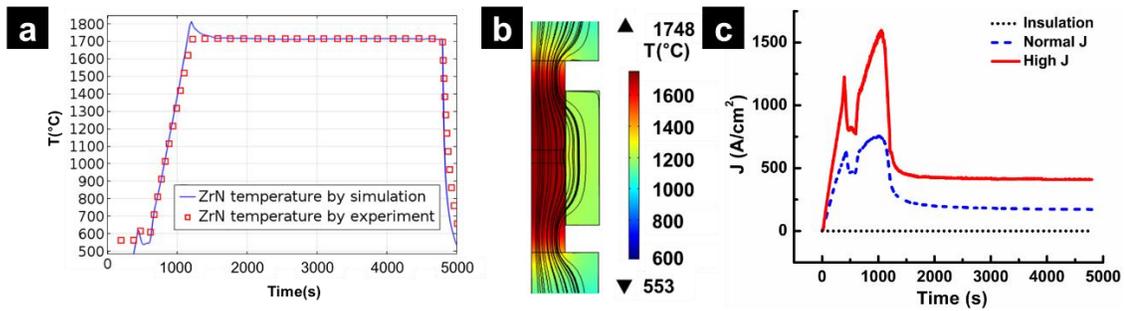

Fig. 2. (a) Experimentally measured and simulated temperatures of ZrN during normal J SPS. (b) FEM simulation map of die set for temperature (color table) and electric current flow (solid line) of normal J mode at the final holding time. (c) Overall local electric current density ($J_{OL}$) change during whole SPS cycle before cooling stage for 3 SPS modes

As shown in Fig. 2(a), the experimentally measured and simulated temperatures by FEM matched well, which indicated that the thermal and electrical parameters in FEM have been well calibrated. Fig. 2(b) shows the FEM simulation map of die set for the temperature and electric current flow of normal J mode, presenting the electric current (black line in Fig. 2(b)) from the top punch was split. Also, the large temperature difference between ZrN sample and die was observed (colored map in Fig. 2(b)), and this rationalize the importance of finding the exact sample temperature instead of the die temperature.

The electric current density of the powders is largely affected by the neck area change during the current-assisted sintering. The summation of the electric current density of all the necks in the powder volume was estimated as overall local electric current density ($J_{OL}$) previously [14].

$$J_{OL} = \frac{I_s}{A_{cross-section}} \times \frac{4\pi R^2}{SZ} \; [A/cm^2] \tag{2}$$

where $I_s$ is electric current passing through the samples (A), $A_{cross-section}$ is cross-sectional area of the full density sample (cm$^2$), $R$ is particle radius (cm), $S$ is average area of contacts between two particles (cm$^2$), and $Z$ is coordination number.

With $I_s$ obtained from the FEM simulation and SPS experiment, $J_{OL}$ was estimated for all 3 SPS modes as shown in Fig. 2(c). The high J mode (red solid line) has the largest $J_{OL}$ during the ramping and the holding stages due to current concentration to the central column of the die set. The normal J mode (blue dash line) has the lower $J_{OL}$ due to the split of the electric current compared with that of the high J mode. We assume that the insulation mode (black dot line) has no electric current flow into ZrN powder.

In the present study, the constitutive equation of the continuum theory of sintering for SPS taking into account the electric current effects based on the electroplasticity theory [5, 14] was used to find the current effects on the sintering. The detailed explanation and derivation were described previously [14]. The constitutive equation of SPS is written as:

$$\dot{\theta} = -\left[\frac{G}{A_0 T}\left(\frac{b}{d}\right)^p \exp\left(\frac{-Q}{RT}\right) + \beta^\omega \left[\int_{t_0}^{t_f} \frac{J_{OL}^2 \lambda}{G} dt\right]^\omega\right] \left(\frac{3\theta}{2}\right)^{\frac{m+1}{2m}} (1-\theta)^{\frac{m-3}{2m}} \left(\frac{\sigma_z}{G}\right)^{\frac{1}{m}} \tag{3}$$

Where $\dot{\theta}$ is the densification rate (*1/s*), $G$ is the shear modulus (MPa), $d$ is the grain size (m), $p$ is the grain size exponent, $T$ is the absolute temperature (K), $b$ is the Burgers vector (m), $Q$ is the creep activation energy (KJ/mol), and $R$ is the gas constant (J/molK), $\theta$ is the porosity, $\sigma_z$ is the Z-axis applied stress (MPa), $m$ is the strain rate sensitivity, $A_0$ is the creep parameter (Pa·s/K), $\beta$ is electric current effect coefficient, $\omega$ is electric current sensitivity, $t_0$, $t_f$ is the starting and final time for SPS respectively (s), and $\lambda$ is electrical resistivity of the powder ($\Omega$cm).

Here, $A_0$ is specified by:

$$A_0 = \frac{k}{A_{Cr} b D_0} \tag{4}$$

Where $k$ is the Boltzmann's constant (J/K), $A_{cr}$ is the material's creep constant, and $D_0$ is the diffusion constant (cm$^2$/s).

With introductions of $A_{TD}$ (1/s) and $A_{ECAD}$ (1/s) indicating the thermal deformability and electric current assisted deformability of the powders [14],

$$A_{TD} = \frac{G}{A_0 T}\left(\frac{b}{d}\right)^p \exp\left(\frac{-Q}{RT}\right) \tag{5}$$

$$A_{ECAD} = \beta^\omega \left[\int_{t_0}^{t_f} \frac{J_{OL}^2 \lambda}{G} dt\right]^\omega \tag{6}$$

the constitutive equation for SPS can be simply written as

$$\dot{\theta} = -[A_{TD} + A_{ECAD}]\left(\frac{3\theta}{2}\right)^{\frac{m+1}{2m}} (1-\theta)^{\frac{m-3}{2m}} \left(\frac{\sigma_z}{G}\right)^{\frac{1}{m}} \tag{7}$$

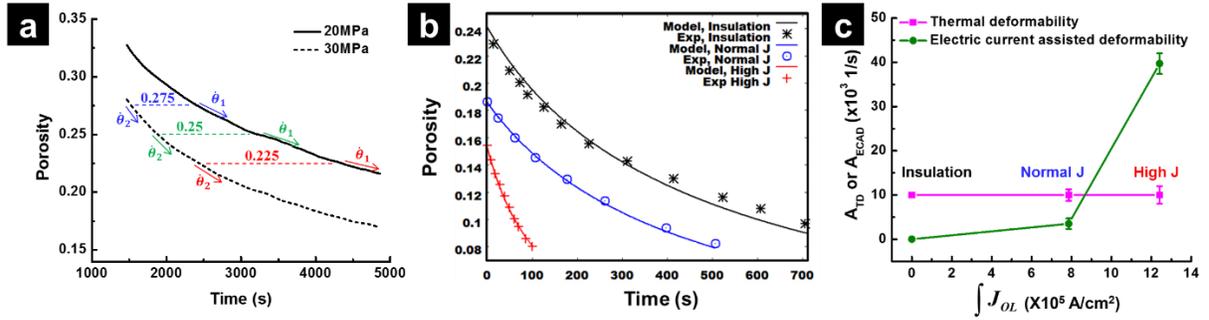

Fig. 3. (a) Porosity evolution of ZrN under 20MPa (solid line) and 30MPa (dash line) at 1500 °C, $\dot{\theta}_1$ and $\dot{\theta}_2$ indicate the densification rate for 20MPa and 30MPa case at specified porosity ($\theta$ = 0.225, 0.25 and 0.275) (b) Experimental porosity evolution curves of ZrN specimens processed by 3 SPS modes fitted by Eq. (3). (c) Comparison of $A_{TD}$ (thermal deformability) and $A_{ECAD}$ (electric current assisted deformability) of ZrN for 3 SPS modes from fitting results.

The relative density lower than 92% during the holding stage was chosen for finding the constitutive parameters for the 3 SPS modes to prevent the possibility of the grain growth effects influencing on the densification mechanism [17]. The porosity evolution (scattered symbols) of ZrN powder (0.08 < $\theta$ < 0.15) for the 3 modes is shown in Fig. 3(b). The high J mode (red cross) shows the fastest densification. At the same time, the densification rate for the insulation mode is the slowest (black asterisk). With the same temperature conditions, the higher the electric current density through ZrN powder is, the faster the densification of the powder is. These results clearly show that the electric current affects the consolidation behavior of ZrN powders.

In the constitutive equation for SPS, 4 parameters ($m$, $A_{TD}$, $\beta$, and $\omega$) in Eq. (3) could be obtained by fitting from the porosity evolution curves. $Q$ is same for all 3 modes because the holding stage with the same temperature was used for the constitutive equation fitting. If we assume that densification mechanism, that is, strain rate sensitivity ($m$), and $A_0$ of the powder are not changed by the electric current [18-20], $m$ and $A_{TD}$ value will be same for all 3 SPS modes. In order to find the effect of current density on $A_{ECAD}$, the following experiment and analysis were done step by step.

At first, $m$ value of used ZrN powders was experimentally obtained by SPS with two different pressure (20 MPa and 30 MPa) at the same temperature (1500 °C) (Fig. 3(a)). If it is assumed that $G$, $A_0$, $Q$, $\beta$, and $d$ for ZrN powders are the same for different pressures ($P_1$ and $P_2$) in Eq. (3) during the sintering, $m$ can be obtained as following [14]:

$$m = \frac{\ln[\frac{P_1}{P_2}]}{\ln[\frac{\dot{\theta}_1}{\dot{\theta}_2}]} \tag{8}$$

Where $\dot{\theta}_1$ and $\dot{\theta}_2$ are the densification rate for 20MPa and 30MPa case at specified porosity ($\theta$ = 0.225, 0.25 and 0.275). Averaged m value of ZrN is 0.504. Larger $m$ value compared with micro-sized ZrN powders ($m$ = 0.22 [14, 21]) can be due to smaller particle size of ZrN powder used in this research [22].

Secondly, with m value from the first step, $A_{TD}$ is only one fitting parameter and is 10042.88 (1/s) by applying constitutive equation for SPS to porosity evolution curve from the insulation mode (Black curve and asterisk symbols in Fig. 3(b)). $A_{ECAD}$ in insulation mode will be zero because no current was applied ($J_{OL}$ =0). All 3 SPS modes will have same $A_{TD}$ value due to the same temperature imposed during SPS.

Lastly, with $m$ and $A_{TD}$ values by previous two steps, and obtained $J_{OL}$ by Eq. (2), $\beta$ and $\omega$ were obtained with Eq. (3) fitting to porosity plots from normal J (Blue curve and circle symbols in Fig. 3(b)) and high J mode (Red curves and cross symbols in Fig. 3(b)), and this gave $A_{ECAD}$ values for each case as shown in Fig. 3(c). The obtained $\omega$ and $\beta$ are 2 and $1.10 \times 10^8$ respectively. $\int J_{OL}$ in Fig. 3(c) indicates the sum of the overall local electric current density ($J_{OL}$) from starting porosity to 92 % of porosity. As shown in Fig. 3(c), $A_{TD}$ is same for all 3 SPS modes, on contrast, $A_{ECAD}$ of 3 SPS modes are different and increased with the amount of the electric current density during the SPS ($\int J_{OL}$). These results show that electric current can affect the plasticity or deformability of the ZrN powder at high temperature.

The m value means the pressure sensitivity of the material in the power law creep behavior [23]. Similarly, this study showed that $\omega$ indeed act as sensitivity value to electric current density As an exponent in the constitutive equation for SPS, $\omega$ plays a critical role in the fitting. In our case, too small or large $\omega$ disturb the good fitting, so obtained $\beta$ and $\omega$ can be unique values for ZrN. This statement will be verified by other materials for later study.

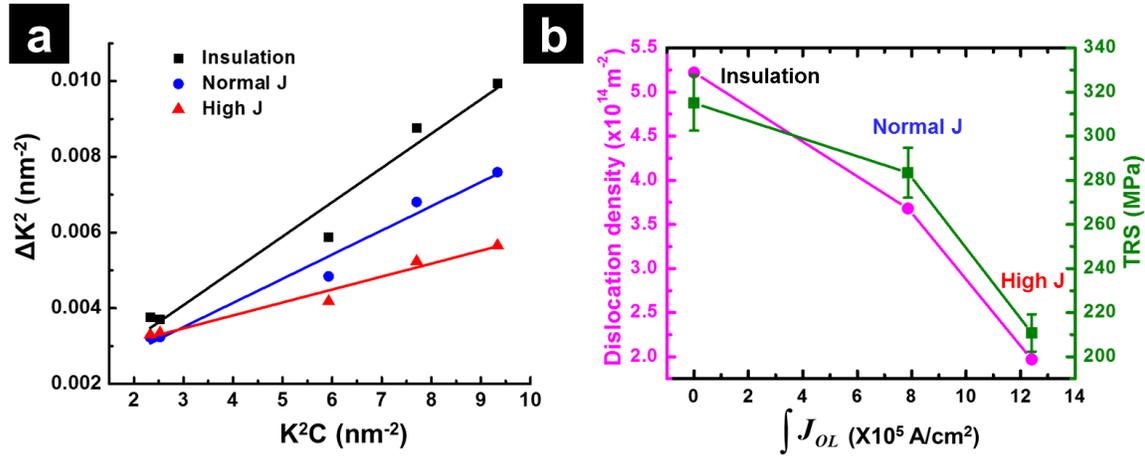

Fig. 4. (a) $\Delta K^2$ vs $K^2C$ plot from the modified Williamson-Hall method, and (b) Relationship among electric current density, dislocation density and Transverse rupture strength (TRS) from 3 SPS modes.

$A_{ECAD}$ in Eq. (3) was created based on an electroplasticity theory which explains the reduced flow stress of the material under electric current is the result of enhanced dislocation motion by local joule heating at defects structure in the lattice [14]. Therefore, we checked if the electric current change the dislocation density of the ZrN pellets applied the different amount of current density.

X-ray diffraction (XRD) (Bruker D-8 diffractometer, MA, USA), utilizing CuKα radiation at room temperature was investigated to ZrN pellets from 3 SPS modes (Fig. S1 in Supplementary Material). The modified Williamson-Hall method was used to estimate the dislocation densities in ZrN pellets [24] (See Supplementary material). It is clear that the slope is decreased with increasing electric current density as shown in Fig. 4(a), which indicates the dislocation density is reduced with larger electric current density. The calculated average dislocation density of ZrN samples from insulation, normal J and high J mode, are $5.22 \times 10^{14}$, $3.68 \times 10^{14}$, and $1.97 \times 10^{14}$ (m$^{-2}$) respectively (Fig. 4(b)). Enhanced dislocation motion by local Joule heating at defects [5] or defect generation [25-27] by electric current can be the reason for the reduction of dislocation density for ZrN pellet from higher current density case. Also, it was observed that with increasing the electric current density, transverse rupture strength (TRS) [28] were decreased too (Fig 4(b)). ZrN samples from 3 SPS modes have a similar relative density (98%) and grain size (~3 μm) (Fig. S2 in Supplementary Material), so TRS results also validate the reduction of dislocation density in ZrN by increasing electric current density. The enhanced dislocation motion decreases the number of dislocation inside ZrN lattice (annihilation), inducing smaller mechanical strength.

In summary, the 3 SPS modes experiments with different electric current density and fitting by constitutive equation for SPS demonstrated that the electric current can influence the deformability of the powders during the sintering. Reduced dislocation density by increasing electric current density based on XRD analysis, also supported by the mechanical strength and microstructural characterization, strongly indicate that electron affects the dislocation motion.

**Acknowledgement**

The support of the U.S. Department of Energy, Materials Sciences Division, under Award No. DE-SC0008581 is gratefully acknowledged.